%% file: elsarticle-template-harv.tex
\documentclass[preprint,12pt,numbers,sort&compress]{elsarticle}

\usepackage{amssymb}
\usepackage{amsmath}
\usepackage{graphicx}
\usepackage{hyperref}
\usepackage{caption}
\usepackage{float}
\usepackage{makecell}
\usepackage{placeins}
\begin{document}

\begin{frontmatter}

\title{Bayesian Localization and Uncertainty Quantification of Trace Species in Two-Dimensional SIMS Imaging}

\author[aff1]{Mengchi~Wang}
\author[aff1]{Binsheu~Shieh}
\author[aff2]{Ichiro~Akai}
\author[aff3]{Masahiro~Hara}
\author[aff4]{Masao~Yoshioka}
\author[aff3]{Takeshi~Hashishin}
\author[aff1]{Toru~Aonishi\corref{cor1}}

\cortext[cor1]{Corresponding author}

\affiliation[aff1]{organization={Graduate School of Frontier Sciences, The University of Tokyo},
            city={Chiba},
            country={Japan}}

\affiliation[aff2]{organization={Institute of Industrial Nanomaterials, Kumamoto University},
            city={Kumamoto},
            country={Japan}}

\affiliation[aff3]{organization={Faculty of Advanced Science and Technology, Kumamoto University},
            city={Kumamoto},
            country={Japan}}

\affiliation[aff4]{organization={Technical Division, Kumamoto University},
            city={Kumamoto},
            country={Japan}}

\begin{abstract}
To enable accurate localization of trace species on material surfaces, we propose a Bayesian framework for analyzing two-dimensional (2D) secondary ion mass spectrometry (SIMS) imaging data. SIMS is widely used in semiconductor manufacturing, materials science, geology, environmental science, and life sciences because of its high sensitivity and excellent elemental and isotopic specificity. However, precise localization remains challenging because of primary ion beam broadening, overlap between neighboring ion distributions, and limited ion counts. The underlying distribution of trace species is modeled as a superposition of two-dimensional Gaussian peaks. To account for the stochastic nature of low-count measurements, the detected ion counts are assumed to follow a Poisson likelihood within a Bayesian framework. Posterior distributions of the peak parameters are estimated using replica-exchange Monte Carlo (REMC), enabling stable inference together with quantitative uncertainty estimation under low-count conditions. The proposed method is first validated using synthetic datasets with known ground truth and is then applied to SIMS measurements of semiconductor samples containing regularly arranged gold (Au) dots with diameters ranging from 0.4 to 2.0~$\mu$m, using scanning electron microscopy (SEM) images as the reference. Optimization of the measurement conditions reduced the relative localization error for 0.4~$\mu$m dots from 5.1\% to 1.9\%. These results demonstrate accurate submicrometer localization with statistically rigorous uncertainty quantification in 2D SIMS imaging.
\end{abstract}


\end{frontmatter}

%% =====================
%% MAIN TEXT
%% =====================

\section{Background}
\input{sections/1_introduction}

\section{Sample preparation and analytical methods}

\subsection{Sample Preparation}
\input{sections/2.1_sample_preparation}

\subsection{Experimental Setup and Ground Truth Definition}
\input{sections/2.2_Experimental_Setup}
\label{sec:2.2_gt}

\subsection{Synthetic Data for Validation}
\input{sections/2.3_synthetic_data.tex}
\label{sec:2.3_synthetic}

\subsection{Bayesian Measurement Framework for SIMS}
\input{sections/2.4_bayesian_framework}

\label{sec:2.4_bayesian}

\section{Results}

\subsection{Sampled Posterior distributions for real SIMS data}
\input{sections/3.1_Uncertainty_quantification}

\subsection{Localization Accuracy under Different Measurement Conditions}
\input{sections/3.2_Localization_accuracy}

\subsection{Validation with Synthetic Data  }
\input{sections/3.3_Synthetic_data}

\subsection{Effect of Signal Accumulation}
\input{sections/3.4_Signal_accumulation}

\section{Discussion}
\input{sections/4_Discussion}

\section{Conclusion}
\input{sections/5_Conclusion}

\clearpage
\appendix
\section{Affine Transformation and Pixel-Size Conversion}
\label{app:affine_transformation}
\input{sections/Appendix}

\section*{Conflicts of interest}
There are no conflicts to declare.

\section*{Data availability}
The data that support the findings of this study are available from the corresponding authors upon reasonable request.

\section*{Acknowledgements}
This work was supported by JST CREST, Grant Number JPMJCR2432.

%% =====================
%% REFERENCES
%% =====================

\bibliography{cas-refs}

\end{document}

%% file: sections/1_introduction.tex
Secondary ion mass spectrometry (SIMS) is a highly sensitive analytical technique that provides spatially resolved elemental and isotopic information. Owing to its ability to perform spatially resolved chemical analysis with high sensitivity, SIMS has been applied extensively in semiconductor device and process analysis,~\cite{vanDerHeide2020,Dhaul2009} materials science and engineered surfaces,~\cite{ImagingSIMS_Chapter,Boulsina2008_ASS,Lian2019_Vacuum} geology and mineralogy,~\cite{Gautier1998} environmental science,~\cite{Milillo2006} and life-science imaging of cells and tissues.~\cite{ToFSIMS_BatteryReview2025,Slodzian1992_BiolCell,GuerquinKern2005_BBA,Green2023_ToFSIMS_LifeSci} In many of these applications, accurate localization of trace species is essential for understanding material properties, interfacial phenomena, and functional performance.~\cite{ToFSIMS_BatteryReview2025,Gautier1998,ImagingSIMS_Chapter,Milillo2006,Slodzian1992_BiolCell,GuerquinKern2005_BBA,Green2023_ToFSIMS_LifeSci} For example, in semiconductor fabrication, even trace amounts of impurities at surfaces or interfaces can significantly affect electrical characteristics, device reliability, and production yield.~\cite{OnizawaCuSiO2,Soden1995}

In SIMS, a focused primary ion beam is raster-scanned across the sample surface to sputter secondary ions, whose mass-to-charge ratios are subsequently analyzed to generate two-dimensional (2D) elemental distribution images.~\cite{Lockyer2024,Slodzian1992_BiolCell,Senoner2010_JAAS}
However, the lateral spatial resolution is fundamentally limited by the finite size of the primary ion beam and by sputtering-induced physical processes such as atomic mixing and surface roughening.~\cite{Boulsina2008_ASS,Lian2019_Vacuum,Senoner2010_JAAS}
Consequently, when characteristic features are smaller than the beam diameter, the measured ion image is spatially blurred.~\cite{Rabasco2022_IJMS_NanoSIMS_Dopamine}
In addition, under low-concentration conditions, only a limited number of ions are detected, resulting in substantial statistical fluctuations.
The combination of spatial blurring and limited ion counts makes accurate localization and rigorous uncertainty quantification particularly challenging.~\cite{Nagata2019,Shieh2026}

In SIMS, deconvolution methods have been proposed to recover spatial resolution and reconstruct the spatial distributions of species~\cite{Lian2019_Vacuum,Boulsina2008_ASS,Li2023,Boulakroune2012,Boulakroune2008_MultiscaleSIMS,Allen1993_MaxEntSIMS,Lee2003_SIMSDeltaLayers}. However, when the objective is to determine the positions of spatially localized trace species rather than reconstruct their spatial distributions, deconvolution is not necessarily required. Other localization methods, such as centroid estimation and least-squares Gaussian fitting, which are widely used for point-source localization in experimental data analysis~\cite{Thompson2002}, are, in principle, applicable to SIMS localization. 
Extending the Bayesian spectroscopy framework originally developed for spectral decomposition in low-count photon measurements~\cite{Shieh2026}
, we propose a Bayesian framework for localizing trace species in low-count two-dimensional (2D) SIMS imaging. Within the proposed framework, the observed SIMS image is represented as a superposition of two-dimensional Gaussian functions, while the detected ion counts are modeled using a Poisson likelihood function. The model parameters are inferred from the posterior distribution conditioned on the observed data, enabling simultaneous localization and uncertainty quantification. Posterior inference is performed using replica-exchange Monte Carlo (REMC), which efficiently samples the posterior distribution through exchanges between replicas at different temperatures.~\cite{Nagata2012,Hukushima1996}

The proposed framework is validated using synthetic datasets with known ground truth (GT) position of trace species and applied to SIMS data of semiconductor samples containing regularly arranged gold (Au) microstructures with diameters ranging from 0.4 to 2.0~$\mu$m. Scanning electron microscopy (SEM) images are used as the reference for quantitative evaluation of localization accuracy. By analyzing the influence of measurement conditions, we demonstrate that optimized acquisition parameters substantially reduce localization errors while providing statistically rigorous uncertainty estimates. These results demonstrate that the proposed framework enables accurate submicrometer localization with reliable uncertainty quantification in low-count 2D SIMS imaging.

%% file: sections/2.1_sample_preparation.tex
\begin{figure}[!htbp]
\centering
\includegraphics[width=0.9\columnwidth]{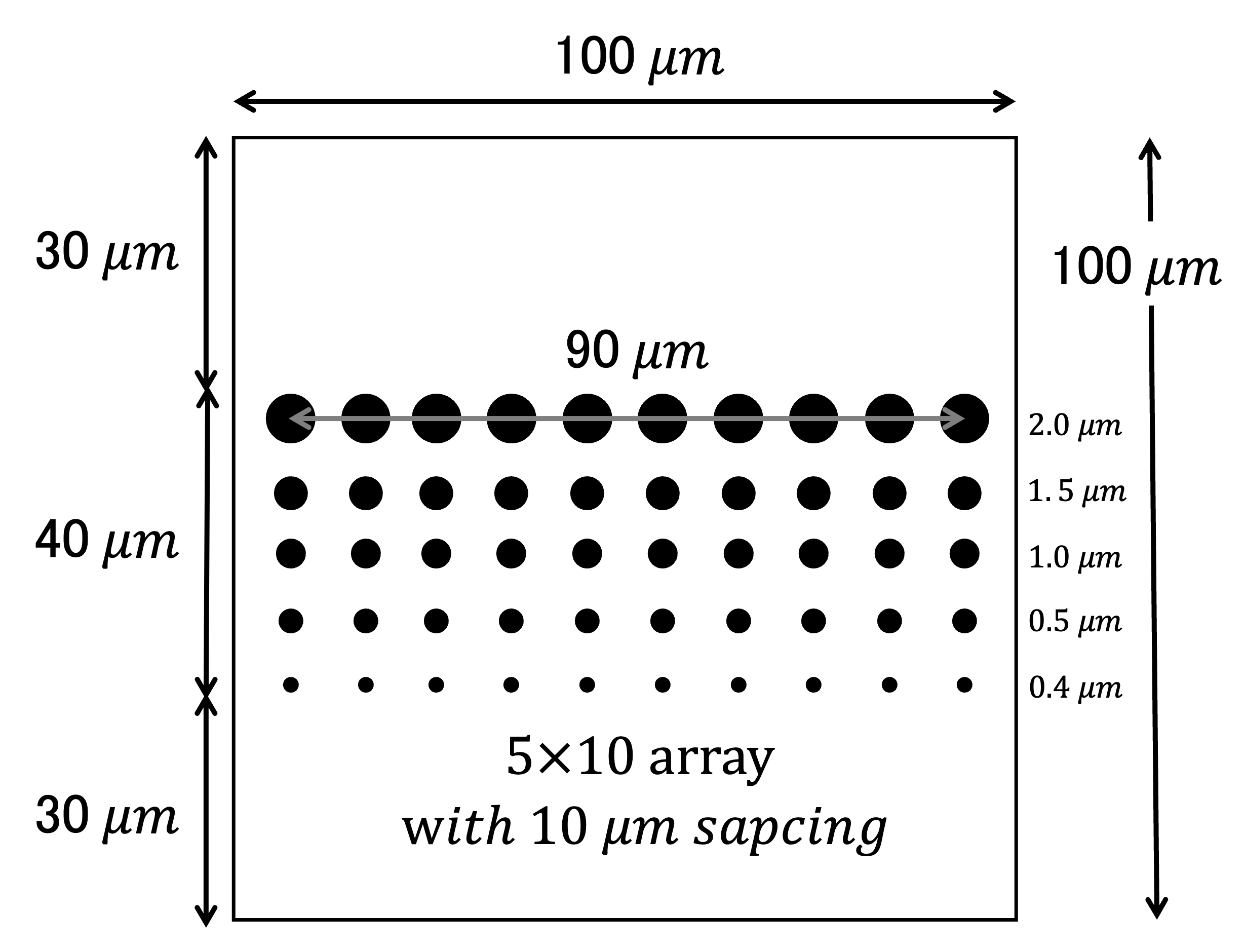}
\caption{
Schematic illustration of the basic Au-dot pattern used in this study.
Au dots with diameters ranging from 0.4 to 2.0~$\mu$m are arranged
in a \(5 \times 10\) array within a
100~$\mu$m $\times$ 100~$\mu$m area, with a spacing of
\(10~\mu\mathrm{m}\) in both the horizontal and vertical directions.
}
\label{fig:sample_schematic}
\end{figure}

A 10~mm $\times$ 10~mm Si substrate containing patterned Au-dot structures
was used in this study. Figure~\ref{fig:sample_schematic} illustrates the
basic Au-dot pattern used for the measurements. In this basic pattern,
Au dots were arranged in a \(5 \times 10\) array within a
100~$\mu$m $\times$ 100~$\mu$m area, with a spacing of
\(10~\mu\mathrm{m}\) in both the horizontal and vertical directions.
Five different dot diameters were prepared: 0.4, 0.5, 1.0, 1.5, and
2.0~$\mu$m.

%% file: sections/2.2_Experimental_Setup.tex
SIMS imaging measurements were performed using a secondary 
ion mass spectrometer (Cameca IMS 7f-Auto) with a Cs$^+$ primary ion beam. 
The detailed measurement conditions are summarized in Table~\ref{tab:measurement_conditions}.

The primary ion beam was operated at a source voltage of 10~kV 
and a sample bias of $-5$~kV, resulting in a net impact energy 
of 15~keV. The beam current was set to 20 and 50~pA. The optical system was configured with primary apertures 
D0 = 3000 and D4 = 200. The secondary apertures consisted of 
a contrast aperture (CA) of 150~$\mu$m and a field aperture (FA) 
of 100~$\mu$m. The entrance and exit slits were set to 598 and 310, 
respectively, with an energy window of 50.2~eV. 
The Dynamic Transfer Optical System (DTOS) mode was enabled. For imaging, raster sizes of $50 \times 50$~$\mu$m$^2$ and 
$60 \times 60$~$\mu$m$^2$ were used with an image resolution of 
$256 \times 256$ pixels and $512 \times 512$ pixels. Each measurement was performed over 
20 acquisition cycles.

\begin{table}[t]
\centering
\caption{SIMS measurement conditions.}
\label{tab:measurement_conditions}
\begin{tabular}{ll}
\hline
\textbf{Category} & \textbf{Condition} \\
\hline

\multicolumn{2}{l}{\textbf{Instrument and primary ion beam conditions}} \\
Instrument & Cameca IMS 7f-Auto \\
Primary ion source & Cs$^+$ \\
Source voltage & 10 kV \\
Sample bias & $-5$ kV \\
Impact energy & 15 keV \\
Beam current & 20 and 50 pA \\
\\
\multicolumn{2}{l}{\textbf{Optical settings}} \\
Primary apertures & D0 = 3000, D4 = 200 \\
Contrast aperture (CA) & 150 $\mu$m \\
Field aperture (FA) & 100 $\mu$m \\
Entrance slit & 598 \\
Exit slit & 310 \\
Energy window & 50.2 eV \\
DTOS mode & Enabled \\
\\
\multicolumn{2}{l}{\textbf{Imaging parameters}} \\
Raster size & $50 \times 50$ and $60 \times 60$ $\mu$m$^2$ \\
Resolution & $256 \times 256$ and $512 \times 512$ pixels \\
Acquisition cycles & 20 \\

\hline
\end{tabular}
\end{table}

\begin{figure}[h]
\centering
\includegraphics[width=0.8\columnwidth]{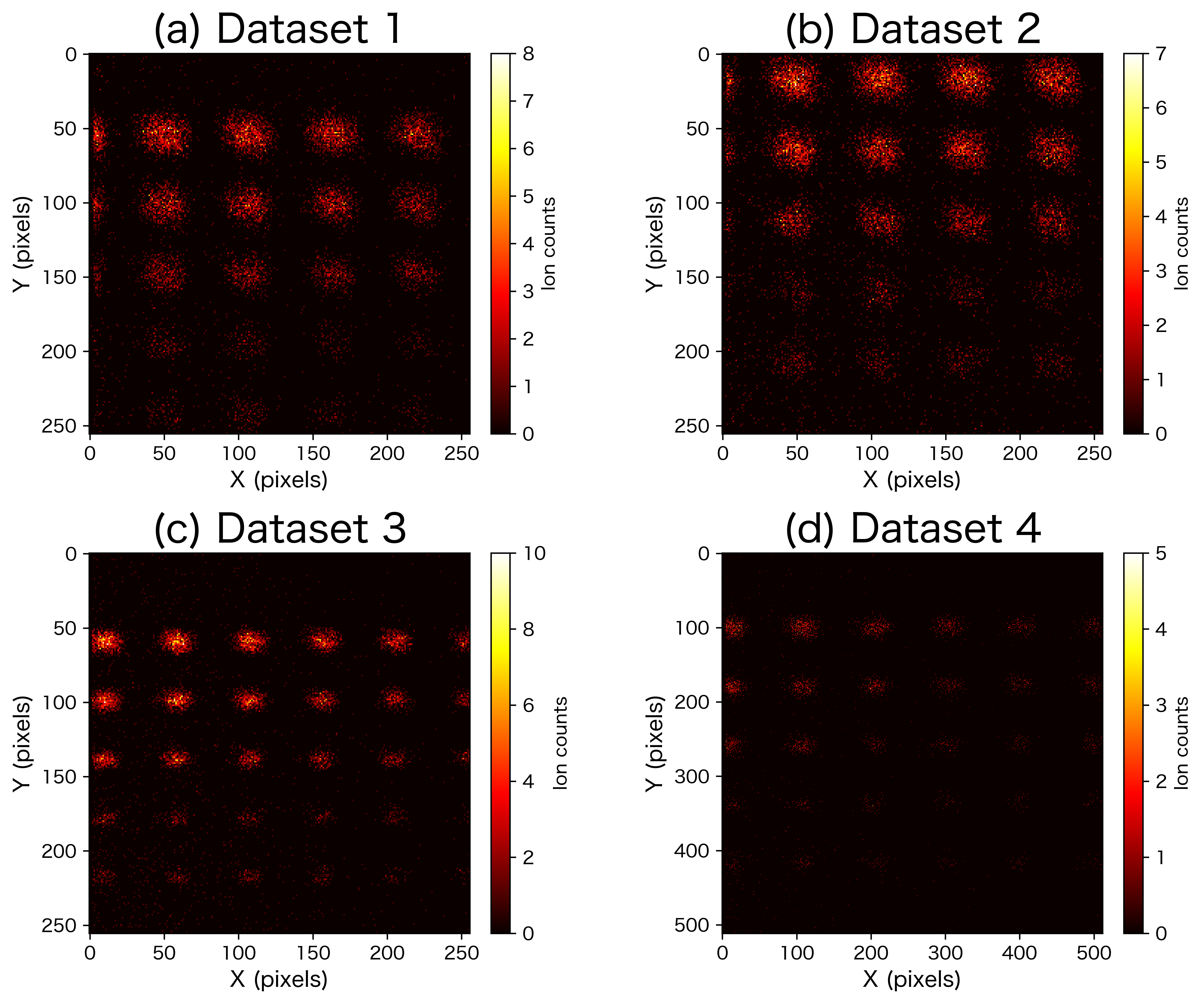}
\caption{
SIMS images acquired under different measurement conditions. 
(a) Dataset 1, (b) Dataset 2, (c) Dataset 3, and (d) Dataset 4. 
Each image shows the ion count distribution of $^{197}$Au obtained during the first of the 20 acquisition cycles for the corresponding dataset. Each image is displayed using an individual color scale.
}
\label{fig:dataset_overview}
\end{figure}

To assess the impact of measurement conditions on Au-dot localization accuracy, 
four datasets (Datasets 1--4) were acquired under varying beam current, 
raster size, measured species, and image resolution 
(Table~\ref{tab:variable_conditions}). 
Fig.~\ref{fig:dataset_overview} presents representative images obtained during the first of the 20 acquisition cycles for each dataset.

\begin{table}[t]
\centering
\caption{Measurement conditions for each dataset.}
\label{tab:variable_conditions}
\small % 或 \footnotesize, \scriptsize

\begin{tabular}{lcccc}
\hline
Parameter & Dataset 1 & Dataset 2 & Dataset 3 & Dataset 4 \\
\hline
Beam current (pA) & 50 & 50 & 20 & 20 \\
Raster size ($\mu$m) & 50 & 60 & 60 & 60 \\
Measured species & $^{29}$Si, $^{197}$Au & $^{28}$Si, $^{197}$Au & $^{28}$Si, $^{197}$Au & $^{28}$Si, $^{197}$Au \\
Resolution & $256 \times 256$ & $256 \times 256$ & $256 \times 256$ & $512 \times 512$ \\
\hline
\end{tabular}
\end{table}

\begin{figure}[H]
\centering
\includegraphics[width=0.8\columnwidth]{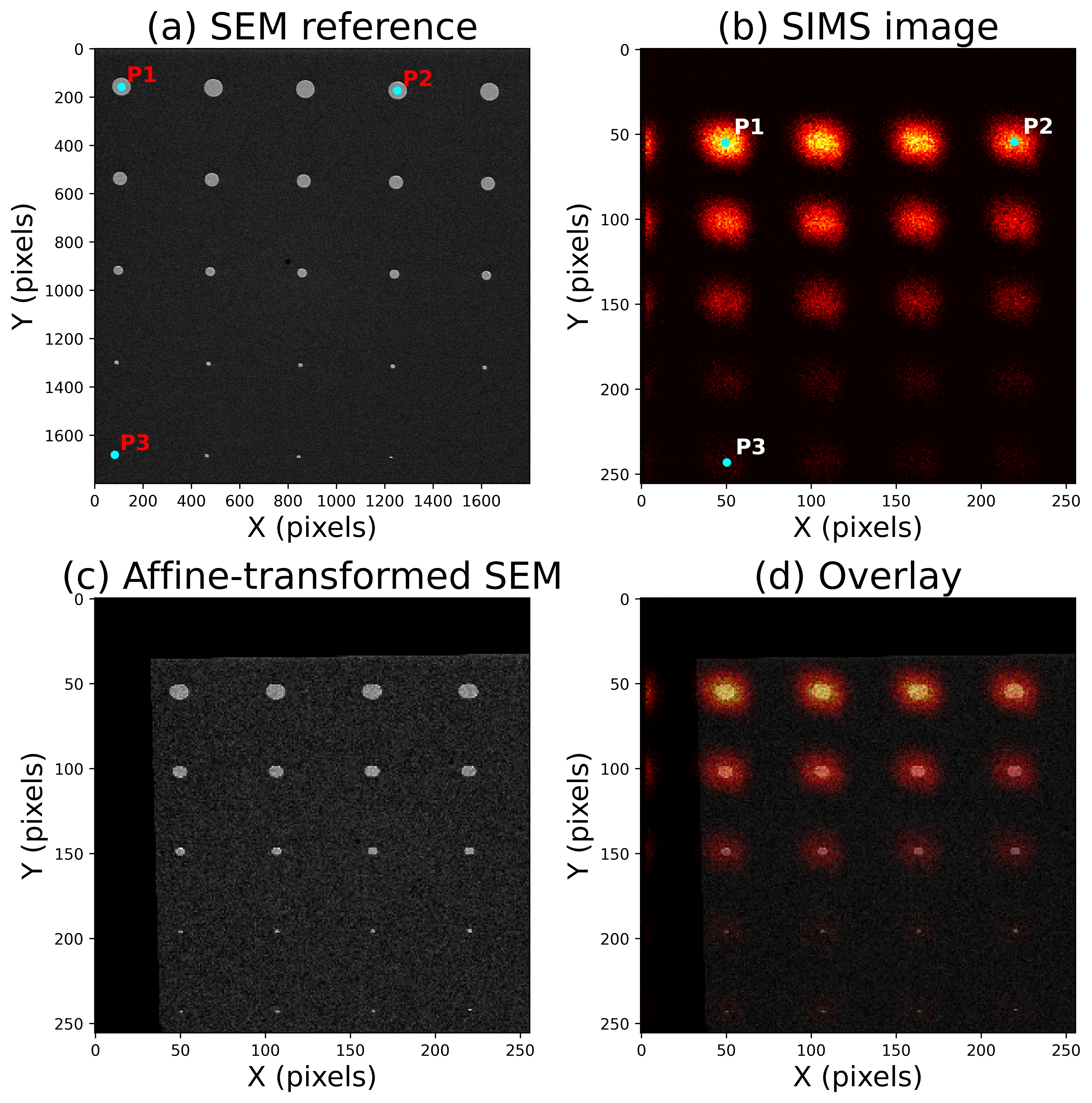}
\caption{
Example of image alignment between the SEM and SIMS images.
(a) SEM image.
(b) SIMS image (Dataset 1) with manually selected corresponding points.
(c) Affine-transformed SEM image. (d) Overlay of the aligned SEM and SIMS images.
}
\label{fig:alignment_example}
\end{figure}

To evaluate localization accuracy, the centers of the Au dots identified in the SEM image were used as the reference positions. As shown in Fig.~\ref{fig:alignment_example}, an affine transformation was estimated from three pairs of corresponding points to register the SEM image to each SIMS image. A detailed description of the affine transformation is provided in~\ref{app:affine_transformation}. Because the horizontal and vertical pixel sizes differ in the SIMS images, the transformed Au dots appear elliptical rather than circular, as shown in Fig.~\ref{fig:alignment_example}(c). The horizontal and vertical pixel sizes vary slightly among the datasets owing to differences in calibration and measurement conditions. These pixel sizes, summarized in 
Table~\ref{tab:pixel_size}, were used to convert the estimated positions and their uncertainties from pixel units to physical length units. After image registration, the Au-dot centers identified in the SEM image were transformed into the coordinate system of each SIMS image. The transformed Au-dot centers, denoted by $(R_x,R_y)$, were used as the reference positions for localization-error evaluation. Because these reference positions were obtained through image registration, they may contain small positional uncertainties arising from the alignment process.

\begin{table}[H]
\centering
\caption{
Physical pixel size  for each dataset.
}
\label{tab:pixel_size}
\begin{tabular}{c c c}
\hline
Dataset &
\makecell{Horizontal pixel size\\$\Delta_h$ (nm/px)} &
\makecell{Vertical pixel size\\$\Delta_v$ (nm/px)} \\
\hline
Dataset 1 & 179.10 & 215.70 \\
Dataset 2 & 174.16 & 214.63 \\
Dataset 3 & 207.66 & 257.19 \\
Dataset 4 & 105.80 & 128.97 \\
\hline
\end{tabular}
\end{table}

%% file: sections/2.3_synthetic_data.tex
To validate the proposed Bayesian framework, we employed the same synthetic datasets as those used in a companion study. \cite{Shieh2026} 
The synthetic datasets consist of two-dimensional images containing multiple localized features arranged on a regular grid. They were generated using the forward model underlying the likelihood function introduced in Section~\ref{sec:2.4_bayesian}

The synthetic datasets were generated to reproduce the signal intensities observed in Dataset~1. The lowest signal intensity was calibrated to match that observed for the 0.4~$\mu$m Au-dot in the first of the 20 acquisition cycles. Signal intensities for the remaining Au-dot diameters were then determined under the assumption that the total signal intensity is proportional to the dot area (i.e., the square of the dot diameter). Consequently, the synthetic datasets comprise five signal-intensity levels corresponding to Au-dot diameters of 0.4, 0.5, 1.0, 1.5, and 2.0~$\mu$m.

Because the synthetic datasets were generated using the same forward model assumed in the proposed Bayesian framework, comparison between the synthetic and experimental results enables indirect assessment of the consistency of the forward model with practical SIMS measurements. Furthermore, because the true parameters are exactly known, the synthetic datasets provide a benchmark for evaluating the intrinsic localization limits of the proposed Bayesian framework independently of uncertainties in the experimental reference positions.

%% file: sections/2.4_bayesian_framework.tex
In this study, we formulate the analysis of SIMS imaging data within a Bayesian framework, in which both the spatial distribution of localized signals and the ion-counting measurement process are modeled probabilistically. This formulation enables simultaneous parameter estimation and uncertainty quantification through posterior inference. The proposed framework is inspired by the Bayesian spectroscopy method of Nagata et al.\cite{Nagata2019}, which was originally developed for decomposing one-dimensional spectra into multiple peaks, but is reformulated here for probabilistic localization in two-dimensional SIMS imaging. The details of the proposed method are as follows.

\subsubsection{Forward model}
The SIMS image formation process is described by the following forward model. The latent spatial signal $S(i,j \mid \Theta)$ at pixel $(i,j)$ is modeled as a superposition of two-dimensional Gaussian spread functions (GPSFs):

\begin{equation}
S(i,j \mid \Theta)
=
\sum_{k=1}^{K}
h_k
\exp
\left[
-\frac{1}{a_k^2}
\left(
(u_i-\mu_k)^2
+
b_k(v_j-\nu_k)^2
+
c_k(u_i-\mu_k)(v_j-\nu_k)
\right)
\right].
\label{eq:gpsf}
\end{equation}

where $\Theta$ denotes the parameter set of the GPSFs,

\begin{equation}
\Theta
=
\{h_k,\mu_k,\nu_k,a_k,b_k,c_k\}_{k=1,\ldots,K}.
\end{equation}

Here, $h_k$ denotes the peak intensity, $(\mu_k,\nu_k)$ denotes the spatial location, $a_k$ denotes the scale parameter, and $b_k$ and $c_k$ characterize the peak anisotropy and orientation.

The observed ion count $y_{ij}$ at pixel $(i,j)$ is modeled as

\begin{equation}
y_{ij}
\sim
\mathrm{Poisson}
\left(
S(i,j \mid \Theta)
\right),
\label{eq:poisson}
\end{equation}

and the likelihood for the entire image
$Y=\{y_{ij}\}$ is given by

\begin{equation}
p(Y \mid \Theta)
=
\prod_{i,j}
\frac{
S(i,j \mid \Theta)^{y_{ij}}
\exp[-S(i,j \mid \Theta)]
}{
y_{ij}!
}.
\label{eq:likelihood}
\end{equation}

\subsubsection{Prior and posterior distributions}

Let $p(\Theta)$ denote the prior distribution representing prior knowledge about the unknown parameter set $\Theta$. Using Bayes' theorem, the posterior distribution of $\Theta$ given the observed image $Y$ is expressed as

\begin{equation}
p(\Theta \mid Y)
\propto
p(Y \mid \Theta)\,p(\Theta)
=
\exp[-N E_N(\Theta)]\,p(\Theta),
\label{eq:posterior}
\end{equation}

where $E_N(\Theta)$ denotes the empirical energy function corresponding to the negative log-likelihood,

\begin{equation}
E_N(\Theta)
=
-\frac{1}{N}
\sum_{i,j}
\left[
y_{ij}\log S(i,j \mid \Theta)
-
S(i,j \mid \Theta)
-
\log(y_{ij}!)
\right].
\label{eq:energy}
\end{equation}

and N is the total number of pixels.

To complete the Bayesian formulation, prior distributions are specified for the model parameters. Independent uniform priors are assumed for all parameters:

\begin{equation}
p(\Theta)
=
\prod_{k=1}^{K}
p(\theta_k),
\label{eq:prior_total}
\end{equation}

where

\begin{equation}
\theta_k
=
(h_k,\mu_k,\nu_k,a_k,b_k,c_k).
\end{equation}

The priors are defined as uniform distributions with the following ranges:

\begin{equation}
0 \le h_k \le 5,\;
0.5 \le a_k \le 20,\;
0.6 \le b_k \le 1.4,
\end{equation}

\begin{equation}
-0.5 \le c_k \le 0.5,\;
5 \le \mu_k \le 35,\;
5 \le \nu_k \le 35.
\label{eq:prior_range}
\end{equation}

These ranges were chosen based on physically plausible values observed in the SIMS measurements and serve to constrain the parameter space during inference.

\subsubsection{Posterior inference using REMC}

Posterior inference is performed using the replica-exchange Monte Carlo (REMC) algorithm, following the framework proposed by Nagata et al.~\cite{Nagata2012,Nagata2019}.  The REMC algorithm is a Markov chain Monte Carlo (MCMC) algorithm designed to efficiently sample from complex posterior distributions~\cite{Hukushima1996}.

To sample from the posterior distribution defined above, the REMC method introduces a set of 
replicated distributions parameterized by an inverse temperature $\beta$. The replicated distributions at different inverse temperatures 
are defined as

\begin{equation}
p_{\beta}(\boldsymbol{\Theta}|\mathbf{Y})
\propto
\exp\left(-\beta N E_N(\boldsymbol{\Theta})\right)
p(\boldsymbol{\Theta}),
\end{equation}

where $\beta \in [0,1]$ controls the smoothness of the 
distribution. When $\beta = 1$, the original posterior 
distribution is recovered, while smaller values of $\beta$ 
correspond to flatter distributions that facilitate global exploration 
of the parameter space. In the REMC framework, a set of replicas 
$\{\boldsymbol{\Theta}_m\}_{m=1,...,M}$ is simulated in parallel, 
each associated with a different inverse temperature $\beta_m$, 
such that $0 \leq \beta_1 < \cdots < \beta_M = 1$.

The algorithm consists of two types of updates.

\begin{description}

\item[Step 1: Local update]
For each replica, the parameter set $\Theta_m$ is updated using the Metropolis--Hastings algorithm based on the corresponding distribution $p_{\beta_m}(\Theta \mid \mathbf{Y})$. For continuous parameters, adaptive step-size tuning based on the Robbins--Monro process is employed to maintain an appropriate acceptance rate and improve sampling efficiency~\cite{Garthwaite2016,Okajima2021}.

\item[Step 2: Exchange between neighboring replicas]
To enhance sampling efficiency, exchanges of configurations between neighboring replicas are proposed. A swap between replicas $m$ and $m+1$ is accepted with probability

\begin{equation}
p(\Theta_m \leftrightarrow \Theta_{m+1})
=
\min(1,\nu),
\label{eq:swap_prob}
\end{equation}

where

\begin{equation}
\begin{aligned}
\nu
&=
\frac{
p_{\beta_m}(\Theta_{m+1}\mid\mathbf{Y})
\,\,
p_{\beta_{m+1}}(\Theta_m\mid\mathbf{Y})
}{
p_{\beta_m}(\Theta_m\mid\mathbf{Y})
\,\,
p_{\beta_{m+1}}(\Theta_{m+1}\mid\mathbf{Y})
}
\\
&=
\exp
\left[
(\beta_{m+1}-\beta_m)
N
\left(
E_N(\Theta_{m+1})
-
E_N(\Theta_m)
\right)
\right].
\end{aligned}
\label{eq:swap_ratio}
\end{equation}

\end{description}

Through these updates, which combine local Metropolis updates with replica exchanges between different temperatures, REMC efficiently explores the posterior distribution. After discarding the burn-in samples, the parameter samples from the replica with inverse temperature $\beta = 1$ are regarded as samples from the posterior distribution $p(\boldsymbol{\theta} \mid \mathbf{Y})$. The maximum a posteriori (MAP) estimate is then obtained by selecting the sampled parameter set with the smallest value of the negative log-posterior:

\begin{equation}
E(\Theta) = -\log p(\Theta \mid Y)
= E_N(\Theta) - \frac{1}{N} \log p(\Theta),
\end{equation}

where $E_N(\Theta)$ is the empirical energy function defined in Eq.~\eqref{eq:energy}, and $p(\Theta)$ is the prior distribution defined in Eq.~\eqref{eq:prior_total}.

\subsubsection{ROI definition and parameter estimation}

\begin{figure}[H]
\centering
\includegraphics[width=0.7\columnwidth]{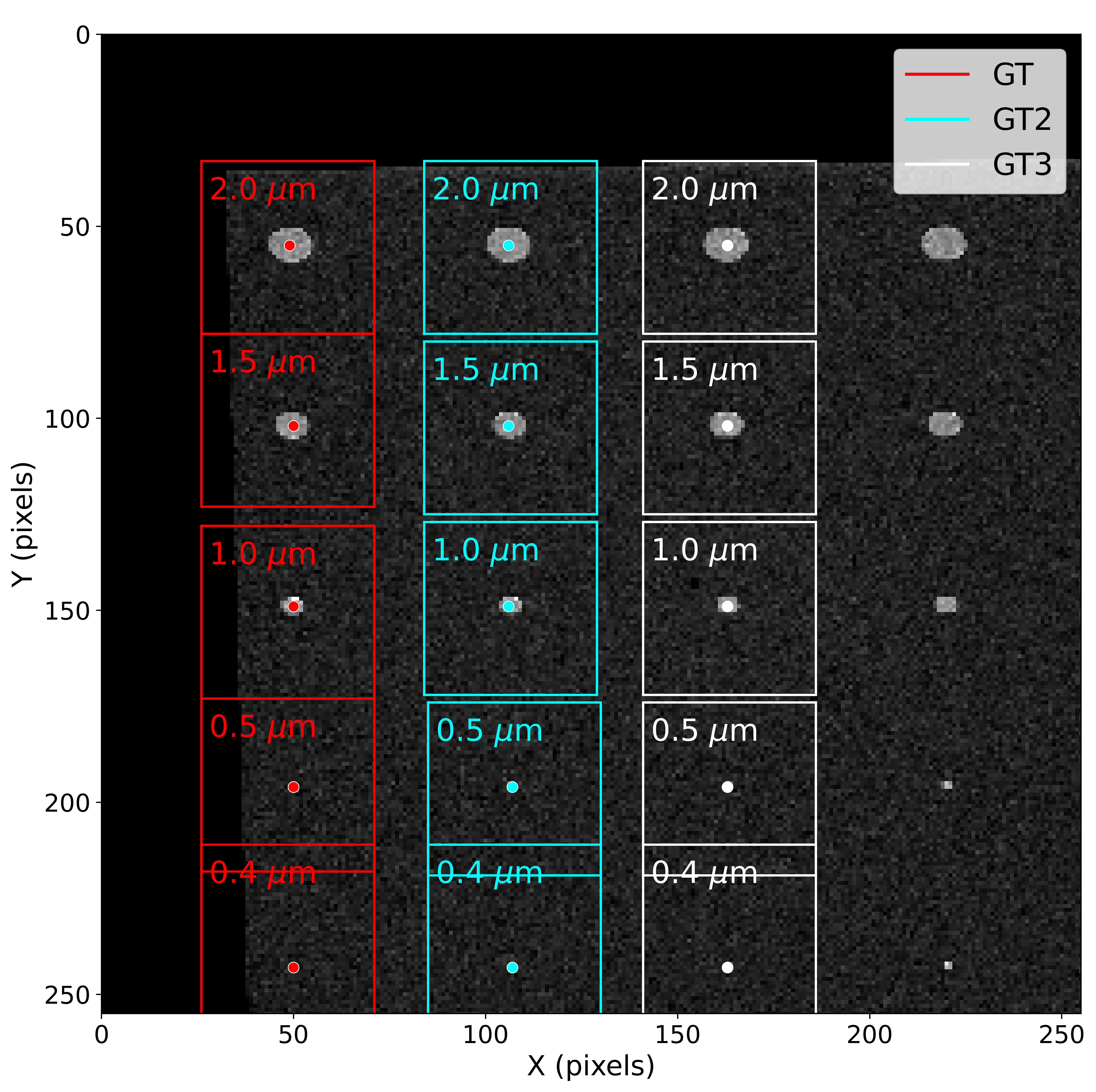}
\caption{
Regions of interest (ROI) segmentation. The observed SIMS image is segmented into ROIs, each containing a single localized signal, and parameter estimation is performed independently for each ROI.
}
\label{fig:roi_schematic}
\end{figure}

For Bayesian inference, the observed SIMS image is partitioned into regions of interest (ROIs) such that each ROI contains a single localized trace species. Fig.~\ref{fig:roi_schematic} shows representative reference position located at different image positions together with their corresponding ROIs. Under this setting, the observed SIMS image in each ROI is represented by a forward model consisting of a single two-dimensional Gaussian component ($K = 1$), and the corresponding model parameters are inferred independently by Bayesian inference.

%% file: sections/3.1_Uncertainty_quantification.tex
We first present representative posterior distributions sampled by REMC from real SIMS data to illustrate the probabilistic behavior of the proposed Bayesian framework. These posterior distributions are then used to quantitatively evaluate the uncertainty associated with the estimated Au-dot position.

\begin{figure}[!htbp]
\centering
\includegraphics[width=0.9\textwidth]{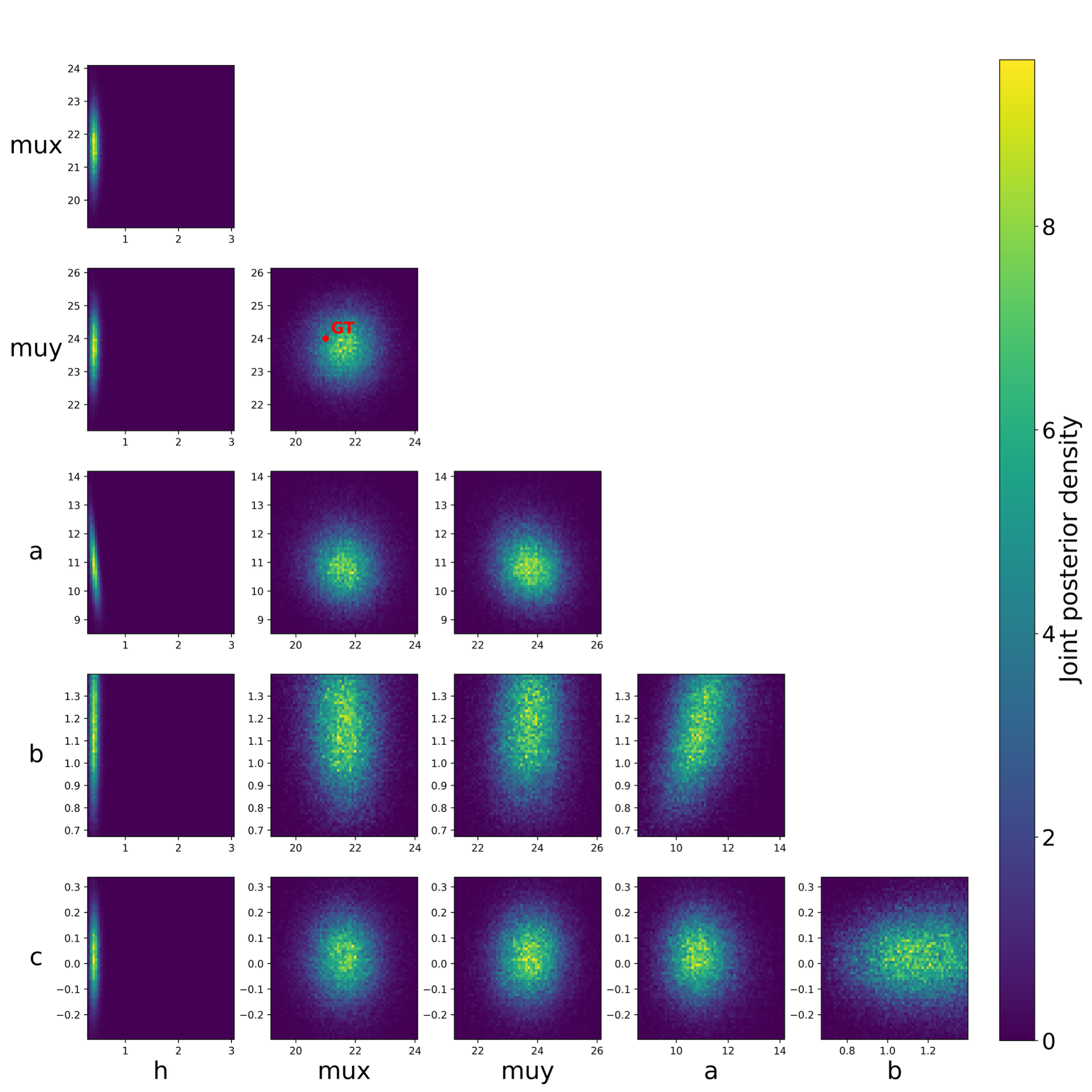}
\caption{
Representative joint posterior distributions for the parameters $(h,\mu_x,\mu_y,a,b,c)$ estimated from the 0.4~$\mu$m condition in Dataset 2. The lower triangular panels show pairwise joint posterior densities obtained from EMC sampling after burn-in removal. The red marker indicates the reference position in the $(\mu_x,\mu_y)$ parameter space.
}
\label{fig:uncertainty_2d}
\end{figure}

To visualize the high-dimensional posterior distribution of the model parameters $(h,\mu_x,\mu_y,a,b,c)$,  Figure~\ref{fig:uncertainty_2d} presents pairwise joint posterior distributions for the first of the 20 acquisition cycles under the 0.4~$\mu$m Au-dot condition in Dataset 2 (Table\ref{tab:variable_conditions}). 
Here, $\mu_x$, $\mu_y$denote the horizontal and vertical coordinates of the center of the Gaussian component, expressed in pixel units, h denotes the peak amplitude, a is the scale parameter controlling the spatial extent of the Gaussian component, and b and c characterize its anisotropy and orientation. The posterior distributions are generally unimodal and concentrated within localized regions of the parameter space. In particular, the joint posterior distribution in the $(h,\mu_x,\mu_y,a,b,c)$ plane forms a compact high-density region centered near the reference position, indicating accurate localization while directly quantifying the associated uncertainty. 

\begin{figure}[!htbp]
\centering
\includegraphics[width=0.9\columnwidth]{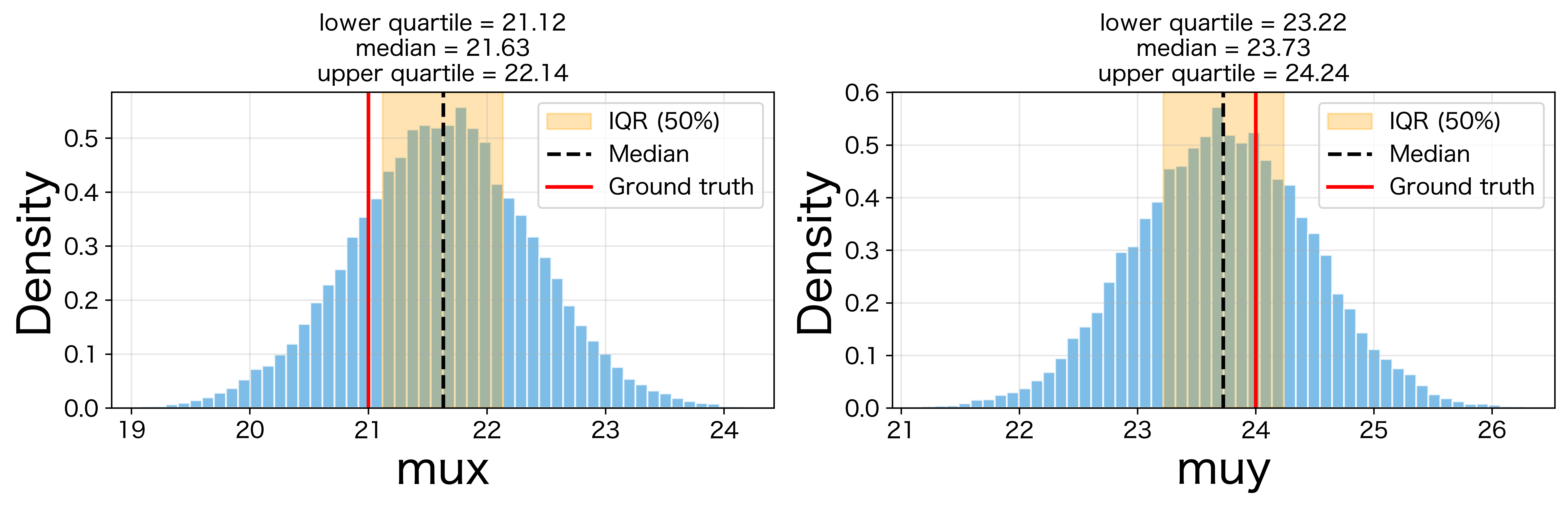}
\caption{
Marginal posterior distributions of the spatial parameters $\mu_x$ and $\mu_y$ or the first acquisition cycle under the 0.4~$\mu$m Au-dot condition in Dataset 2. The shaded regions indicate the IQR, and the red vertical lines denote the reference positions.
}
\label{fig:posterior_example}
\end{figure}

For a more quantitative evaluation of localization uncertainty, Figure 6 presents the marginal posterior distributions of the spatial parameters $\mu_x$ and $\mu_y$, derived from the same posterior samples used to construct the pairwise joint posterior distributions in Figure~\ref{fig:uncertainty_2d}. The shaded regions indicate the interquartile range (IQR), corresponding to the central 50\% probability mass of each posterior distribution. For the example shown in Figure~\ref{fig:posterior_example}, the IQR was 1.02 pixels (177.64~nm) in the horizontal direction and 1.02 pixels (218.92~nm) in the vertical direction. The corresponding half-IQR values, measured from the posterior median, were 0.51 pixels (88.82~nm) and 0.51 pixels (109.46~nm), respectively. These values quantify the localization uncertainty of the estimated Au-dot position.

As shown in Figure~\ref{fig:posterior_example}, the reference position lies within the IQR of the posterior distribution for $\mu_y$ and close to the IQR boundary for $\mu_x$. This result indicates that the posterior distributions provide a reasonable quantitative description of the localization uncertainty for this example.

\newpage

%% file: sections/3.2_Localization_accuracy.tex
To quantitatively evaluate the localization performance of the proposed Bayesian framework, localization errors were evaluated for the first acquisition cycle under each of the four measurement conditions (Datasets 1–4; Table\ref{tab:variable_conditions}). The estimated position of each Au dot was obtained from the location parameters $(\mu_k,\nu_k)$ of the maximum a posteriori (MAP) estimate.

The localization error was defined as the Euclidean distance between the estimated position and the reference position,

\begin{equation}
d=
\sqrt{
(\mu_k-R_x)^2+
(\nu_k-R_y)^2
},
\label{eq:localization_error}
\end{equation}

where $(R_x,R_y)$ denotes the reference positions described in Section~\ref{sec:2.2_gt}. For comparison among different experimental conditions, the relative error was additionally defined by normalizing the localization error with respect to the nominal dot spacing of 10~$\mu$m.

\begin{figure}[!htbp]
\centering
\includegraphics[width=0.6\columnwidth]{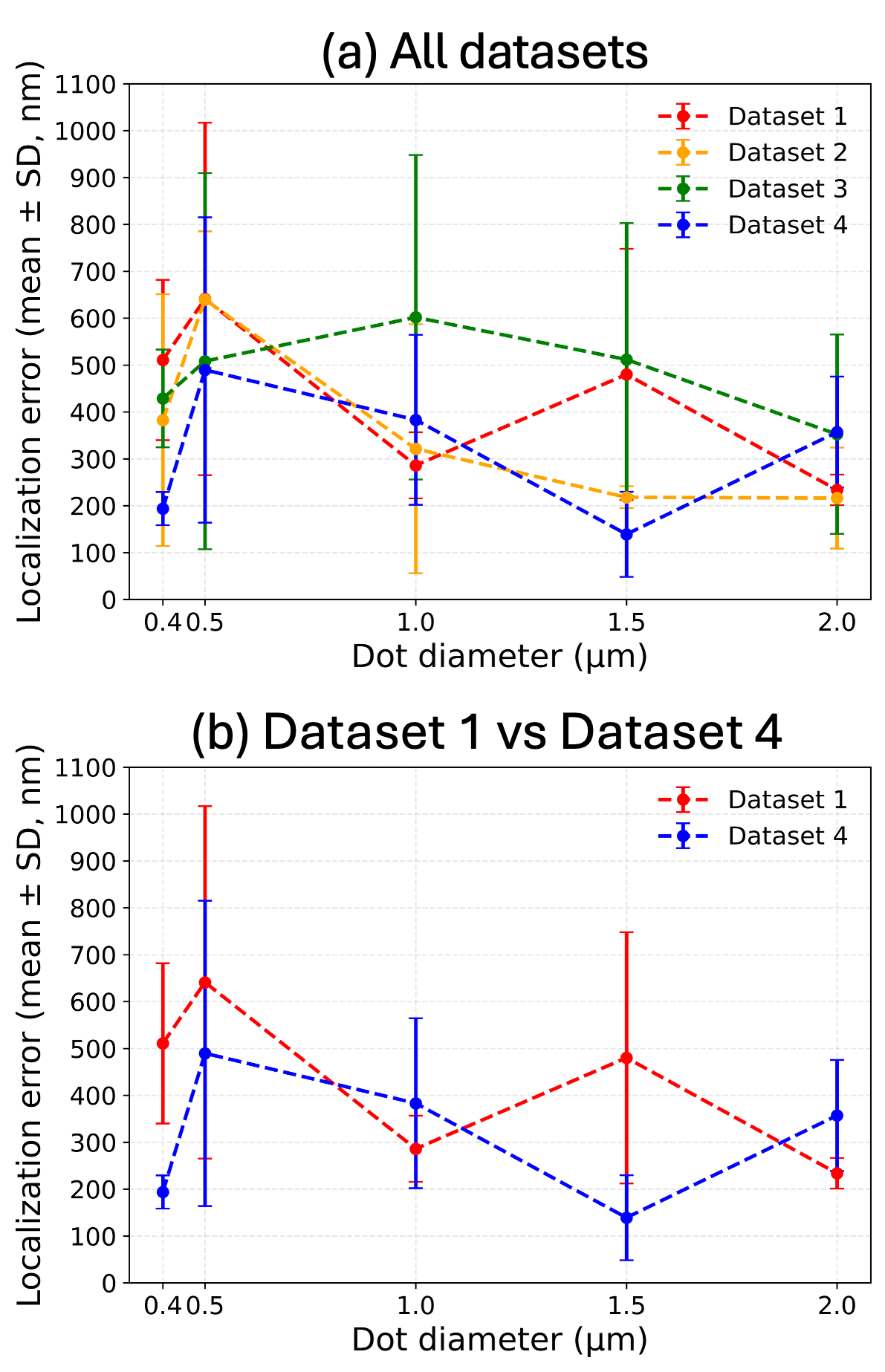}
\caption{
Localization error as a function of Au dot diameter.
(a) Localization errors for the first acquisition cycle under the four measurement conditions (Datasets 1–4).
(b) Direct comparison between Dataset 1 and Dataset 4.
}

\label{fig:measurement_condition}
\end{figure}

Figure~\ref{fig:measurement_condition}(a) shows the localization error as a function of Au-dot diameter for the four measurement conditions (Datasets 1–4), and Table\ref{tab:localization_error} summarizes the corresponding numerical values. Each curve corresponds to one dataset. The localization error tended to decrease as the Au-dot diameter increased. Although this trend was generally consistent across all datasets, the absolute localization error depended strongly on the measurement conditions. Most relative localization errors were approximately 6\% or lower, indicating sub-micrometer localization accuracy under all measurement conditions.

Figure~\ref{fig:measurement_condition}(b)  compares the localization errors obtained for Datasets 1 and 4, which were acquired with different SIMS image resolutions. For three of the five Au-dot diameters, Dataset 4 exhibited lower localization errors than Dataset 1. In particular, for the smallest 0.4~µm Au dots, the mean localization error decreased substantially from 511.0~nm in Dataset 1 to 193.8~nm in Dataset 4. This result suggests that higher SIMS spatial resolution contributes to improved localization accuracy, particularly for weak and spatially confined signals.

\FloatBarrier

\begin{table}[!htbp]
\centering
\caption{
Localization error under different measurement conditions as a function of the Au dot diameter, $d$.
Values in parentheses indicate the relative error normalized by the nominal dot spacing (10~$\mu$m).
}
\label{tab:localization_error}
\footnotesize
\setlength{\tabcolsep}{1.5pt}
\renewcommand{\arraystretch}{1.0}
\begin{tabular}{cccccc}
\hline
d ($\mu$m)
& Dataset 1
& Dataset 2
& Dataset 3
& Dataset 4 \\
\hline
0.4 &
511.0 $\pm$ 171.0 (5.1\%) &
382.5 $\pm$ 268.4 (3.8\%) &
428.9 $\pm$ 104.3 (4.3\%) &
193.8 $\pm$ 35.3 (1.9\%) \\
0.5 &
641.1 $\pm$ 376.1 (6.4\%) &
639.6 $\pm$ 145.5 (6.4\%) &
508.3 $\pm$ 401.2 (5.1\%) &
489.7 $\pm$ 325.7 (4.9\%) \\
1.0 &
286.1 $\pm$ 70.7 (2.9\%) &
321.4 $\pm$ 265.7 (3.2\%) &
601.8 $\pm$ 346.0 (6.0\%) &
383.1 $\pm$ 181.1 (3.8\%) \\
1.5 &
480.1 $\pm$ 268.0 (4.8\%) &
217.9 $\pm$ 23.3 (2.2\%) &
511.8 $\pm$ 291.0 (5.1\%) &
139.1 $\pm$ 90.7 (1.4\%) \\
2.0 &
233.7 $\pm$ 32.5 (2.3\%) &
216.3 $\pm$ 107.8 (2.2\%) &
352.4 $\pm$ 212.7 (3.5\%) &
357.1 $\pm$ 118.5 (3.6\%) \\
\hline
\end{tabular}
\end{table}

%% file: sections/3.3_Synthetic_data.tex
To evaluate the intrinsic localization limits of the proposed Bayesian framework and to assess the consistency of the proposed forward model with practical SIMS measurements, the proposed method was applied to the synthetic datasets described in Section~\ref{sec:2.3_synthetic}, for which the reference positions are exactly known.

\begin{figure}[!h]
\centering
\includegraphics[width=0.9\columnwidth]{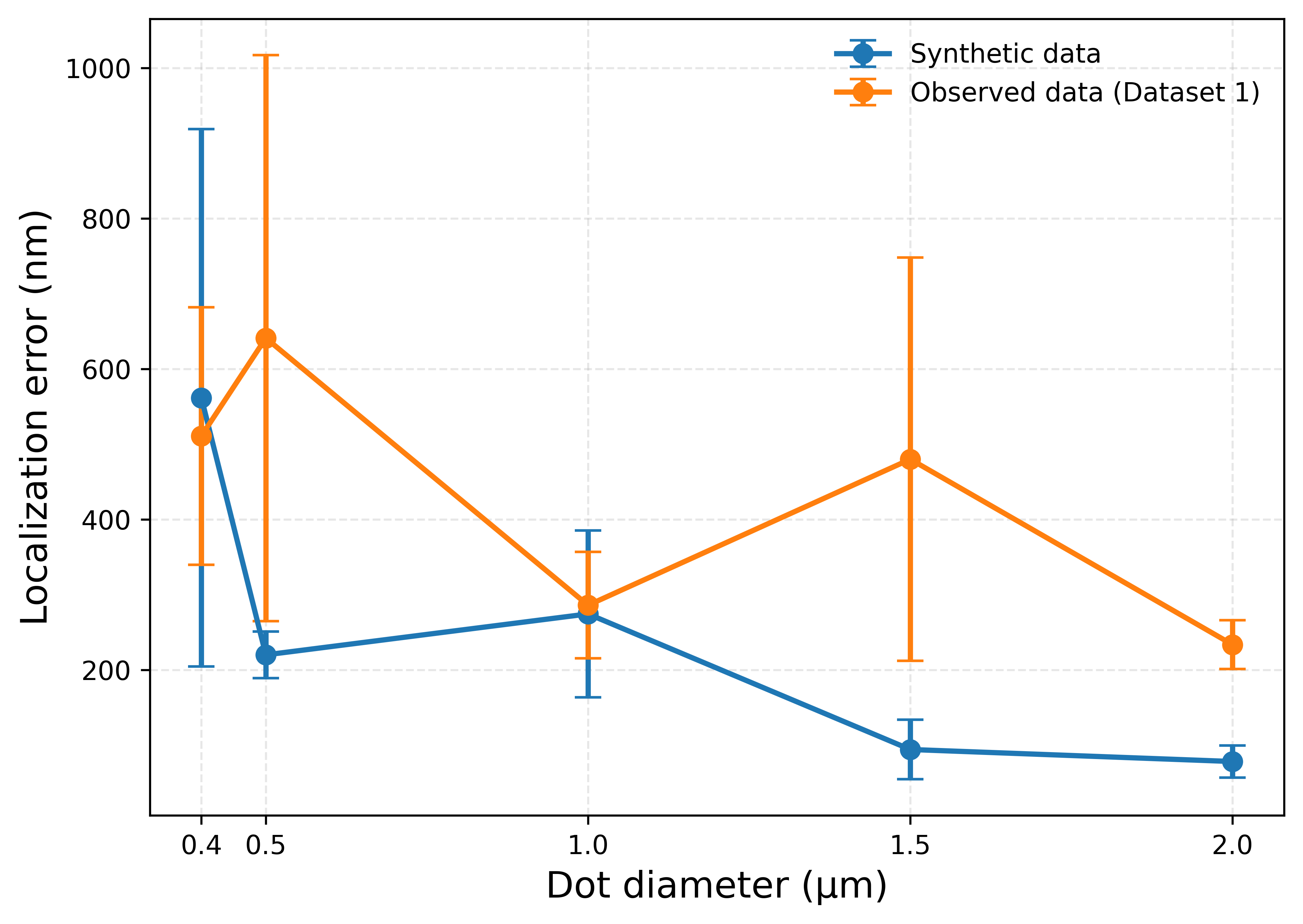}
\caption{
Comparison of localization errors between the synthetic dataset and Dataset 1 as a function of dot diameter.}
\label{fig:synthetic_example}
\end{figure}

Figure \ref{fig:synthetic_example}  compares the localization errors obtained from the synthetic and experimental datasets (Dataset 1) as a function of Au-dot diameter, and Table~\ref{tab:synthetic_comparison} summarizes the corresponding numerical values. For the 0.4~µm and 1.0~µm Au-dot diameters, the mean localization errors obtained from the synthetic and experimental datasets were nearly identical. For the remaining Au-dot diameters, the localization errors obtained from the experimental data were mostly 1–2 pixels larger than those obtained from the synthetic data. Although a small difference in absolute localization error was observed between the two datasets, both exhibited qualitatively similar trends across the investigated Au-dot diameters. This qualitative agreement suggests that the proposed forward model provides a reasonable description of the SIMS signal generation process. 

\begin{table}[H]
\centering
\caption{Comparison of localization errors between synthetic and experimental data (Dataset~1).}
\label{tab:synthetic_comparison}

\begin{tabular}{ccc}
\hline
$d$ ($\mu$m) & Synthetic & Dataset 1 \\
\hline
0.4 & 561.7 $\pm$ 357.3 (5.6\%) & 511.0 $\pm$ 171.0 (5.1\%) \\
0.5 & 220.1 $\pm$ 30.8 (2.2\%) & 641.1 $\pm$ 376.1 (6.4\%) \\
1.0 & 274.5 $\pm$ 111.0 (2.7\%) & 286.1 $\pm$ 70.7 (2.9\%) \\
1.5 & 94.3 $\pm$ 39.6 (0.9\%) & 480.1 $\pm$ 268.0 (4.8\%) \\
2.0 & 78.4 $\pm$ 21.4 (0.8\%) & 233.7 $\pm$ 32.5 (2.3\%) \\
\hline
\end{tabular}
\end{table}

%% file: sections/3.4_Signal_accumulation.tex
\begin{figure}[H]
\centering
\includegraphics[width=0.9\columnwidth]{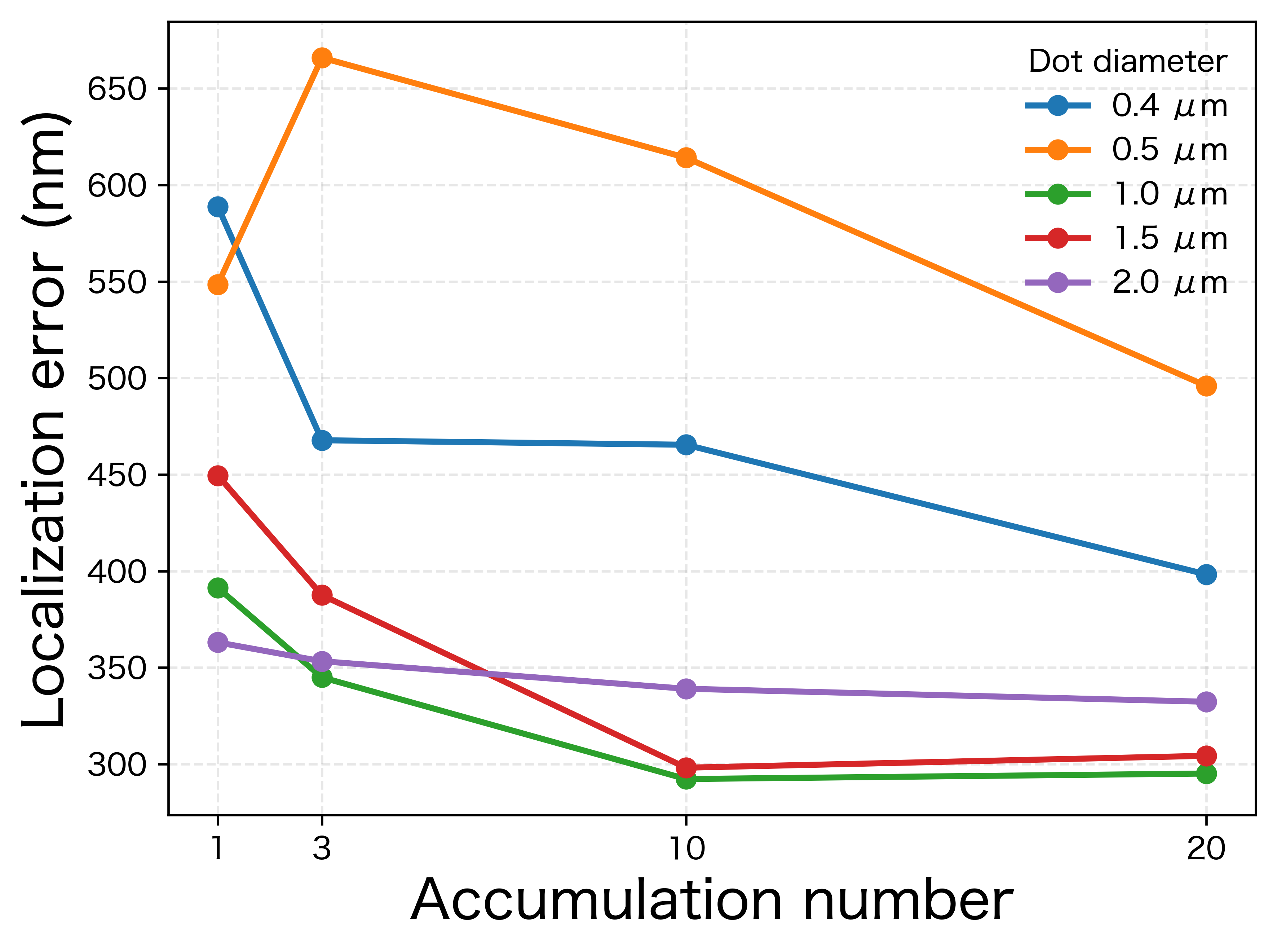}
\caption{Localization error as a function of accumulation number for Dataset 1. Each curve corresponds to a different dot diameter. 
}
\label{fig:accumulation}
\end{figure}

To investigate the effect of signal accumulation on localization accuracy, the number of accumulated images was varied from 1 to 20 for each Au-dot diameter in Dataset 1. Here, Acc.~1 corresponds to the first acquisition cycle alone, whereas
Acc.~3, Acc.~10, and Acc.~20 denote images obtained by accumulating the
first 3, 10, and all 20 acquisition cycles, respectively. Figure~\ref{fig:accumulation}  shows the localization error as a function of the number of accumulated images for different Au-dot diameters, and Table 6 summarizes the corresponding numerical values. Overall, the localization error tended to decrease as the number of accumulated images increased. An exception was observed for the 0.5~µm Au-dot condition, where the localization error increased between accumulation numbers 1 and 3 before decreasing with further accumulation. Moreover, even after averaging 20 measurements, the localization error for the 0.5~µm Au-dot remained larger than that for the 0.4~µm Au-dot despite the larger Au-dot diameter.

\begin{table}[H]
\centering
\caption{
Localization error (nm) as a function of accumulation number for Dataset~1.
}
\label{tab:accumulation_dataset1}

\begin{tabular}{c c c c c}
\hline
Dot diameter ($\mu$m)
& Acc. 1
& Acc. 3
& Acc. 10
& Acc. 20 \\
\hline
0.4 & 588.8 & 467.8 & 465.4 & 398.3 \\
0.5 & 548.5 & 665.9 & 614.1 & 495.9 \\
1.0 & 391.3 & 344.9 & 292.3 & 295.1 \\
1.5 & 449.5 & 387.5 & 298.1 & 304.3 \\
2.0 & 363.0 & 353.3 & 339.0 & 332.3 \\
\hline
\end{tabular}

\end{table}

%% file: sections/4_Discussion.tex
\subsection{Bayesian localization and uncertainty quantification}

The present study formulates localization in low-count SIMS imaging as a Bayesian inference problem by explicitly modeling the forward imaging process, in which localized species are observed as spatially spread ion signals governed by stochastic ion-counting statistics. Unlike conventional deterministic localization methods, the proposed framework estimates not only the most probable location of each localized species but also its posterior distribution, as demonstrated in Figure~\ref{fig:uncertainty_2d} and Figure~\ref{fig:posterior_example}, thereby enabling simultaneous localization and quantitative uncertainty assessment. This probabilistic formulation provides a principled framework for uncertainty-aware localization in SIMS imaging, where localization accuracy is fundamentally limited by the spatial spreading of SIMS signals and statistical fluctuations arising from limited ion counts.

\subsection{Influence of measurement conditions}

The experimental results in Figure~\ref{fig:measurement_condition} demonstrate that localization accuracy depends strongly on both the ion-count level and the measurement conditions. Larger Au dots produce higher ion counts, resulting in narrower posterior distributions and more accurate localization. Comparison between Datasets 1 and 4 further indicates that higher SIMS image resolution substantially improves localization accuracy. The superior localization performance of Dataset 4 implies that finer spatial sampling is especially beneficial for weak and spatially confined signals. These findings indicate that the proposed Bayesian framework can serve not only as a localization method but also as a quantitative tool for evaluating and optimizing SIMS measurement conditions from the viewpoint of localization performance.

\subsection{Validation using synthetic data}

The synthetic-data analysis in Figure \ref{fig:synthetic_example} provides two important findings. First, the qualitative agreement between the synthetic and experimental localization trends shown in Figure \ref{fig:synthetic_example} provides additional support for the validity of the proposed forward model, suggesting that it captures the essential characteristics of the SIMS signal generation process in practical measurements. Second, the consistently larger localization errors observed for the experimental data suggest that uncertainties in the experimental reference positions contribute to the measured localization errors. This interpretation is further supported by the 0.5~µm Au-dot condition, which exhibited relatively large localization errors throughout multiple analyses, including the signal-accumulation study shown in Figure~\ref{fig:accumulation}, despite its larger diameter. This observation suggests that the GT position determined for this Au-dot may contain a larger positional uncertainty than those of the other Au-dot conditions.

\subsection{Limitations and future extensions}

The present study considered the simplest case in which each ROI contains a single localized species represented by a single two-dimensional Gaussian function (K = 1). Although this assumption enabled quantitative evaluation of localization accuracy and uncertainty using isolated Au-dot structures, the proposed forward model represents the observed SIMS image as a superposition of two-dimensional Gaussian functions and can therefore be readily extended to partially overlapping signals and more complex spatial distributions by increasing the number of Gaussian components. Furthermore, because Bayesian posterior inference is performed using REMC, the Bayesian free energy can be evaluated, enabling Bayesian model selection to determine the appropriate number of Gaussian components. Consequently, the proposed framework provides a foundation for Bayesian analysis of increasingly complex SIMS imaging data involving multiple localized species and complex impurity distributions. 

%% file: sections/5_Conclusion.tex
In this study, we proposed a Bayesian framework for localizing trace species in low-count two-dimensional SIMS imaging. The proposed framework explicitly models the forward imaging process, in which localized species are observed as spatially spread ion signals governed by stochastic ion-counting statistics. The posterior distribution of the model parameters is inferred using REMC, from which both the locations of trace species and their associated uncertainties are quantitatively estimated. Application of the proposed framework to semiconductor SIMS measurements demonstrated accurate submicrometer localization over a wide range of Au-dot diameters and imaging conditions. The inferred posterior distributions enabled quantitative evaluation of localization uncertainty, while comparison with synthetic datasets generated from the same forward model provided additional support for the validity of the proposed forward model and enabled evaluation of the intrinsic localization limits of the proposed Bayesian framework. More generally, the present study demonstrates that Bayesian probabilistic modeling provides a practical framework for uncertainty-aware localization in SIMS imaging. Because the proposed forward model represents SIMS images as a superposition of two-dimensional Gaussian components, the same Bayesian framework can, in principle, be extended to more complex SIMS images containing overlapping signals and multiple localized species through Bayesian model selection. The proposed methodology therefore provides a foundation for quantitative and uncertainty-aware analysis of increasingly complex SIMS imaging data.

%% file: sections/Appendix.tex
The affine transformation used to align the SEM image with the SIMS image
is defined as

\begin{equation}
\begin{bmatrix}
x' \\
y'
\end{bmatrix}
=
\begin{bmatrix}
a & b \\
c & d
\end{bmatrix}
\begin{bmatrix}
x \\
y
\end{bmatrix}
+
\begin{bmatrix}
t_1 \\
t_2
\end{bmatrix},
\label{eq:affine_transformation}
\end{equation}

where $[x',y']^{T}\in\mathbb{R}^{2}$ denotes a coordinate in the SIMS
image, and $[x,y]^{T}\in\mathbb{R}^{2}$ denotes the corresponding
coordinate in the SEM image. The matrix with elements $a$, $b$, $c$, and
$d$ represents the linear transformation, including rotation, scaling, and
distortion, while $[t_1,t_2]^{T}\in\mathbb{R}^{2}$ denotes the translation
vector. The transformation parameters were estimated from three pairs of
manually selected corresponding points between the SEM and SIMS images.

One horizontal and one vertical pixel in the SIMS image correspond to the
vectors $[\delta x_h,\delta y_h]^T$ and
$[\delta x_v,\delta y_v]^T$, respectively, in the SEM image coordinates.
These vectors satisfy

\begin{equation}
\begin{bmatrix}
1 \\
0
\end{bmatrix}
=
\begin{bmatrix}
a & b \\
c & d
\end{bmatrix}
\begin{bmatrix}
\delta x_h \\
\delta y_h
\end{bmatrix},
\qquad
\begin{bmatrix}
0 \\
1
\end{bmatrix}
=
\begin{bmatrix}
a & b \\
c & d
\end{bmatrix}
\begin{bmatrix}
\delta x_v \\
\delta y_v
\end{bmatrix}.
\label{eq:pixel_vectors}
\end{equation}

Therefore, the corresponding lengths in the SEM image, $s_h$ and $s_v$,
are given by

\begin{equation}
s_h
=
\sqrt{(\delta x_h)^2+(\delta y_h)^2}
=
\frac{\sqrt{d^2+c^2}}{|ad-bc|},
\qquad
s_v
=
\sqrt{(\delta x_v)^2+(\delta y_v)^2}
=
\frac{\sqrt{a^2+b^2}}{|ad-bc|}.
\label{eq:pixel_scaling_factors}
\end{equation}

The physical pixel sizes of the SIMS image in the horizontal and vertical
directions, $\Delta_h$ and $\Delta_v$, are then calculated as

\begin{equation}
\Delta_h=\Delta_0 s_h,
\qquad
\Delta_v=\Delta_0 s_v,
\label{eq:physical_pixel_size}
\end{equation}

where $\Delta_0$ denotes the physical pixel size of the SEM image. Since
the SEM image has isotropic pixels, the physical pixel size is identical
in the horizontal and vertical directions.